# DMR-based Technique for Fault Tolerant AES S-box Architecture

Mahdi Taheri[1], Saeideh Sheikhpour[1], Mohammad Saeed Ansari[2] and Ali Mahani[1]
[1]Department of Electrical Engineering, Shahid Bahonar University, Kerman, Iran
[2]Eideticom Computational Storage, Calgary, AB, Canada

*Abstract*— This paper presents a high-throughput fault-resilient hardware implementation of AES S-box, called HFS-box. We propose deep pipelining S-box in the gate level in which a novel DMR technique is used for fault correction. Proposed fault-resilient technique is based on fault correction in DMR implementation (FC-DMR) of each S-box's combined with a temporal redundancy technique. If a transient natural or even malicious fault(s) in each pipeline stage is detected, the corresponding error signal become high and as a result the control unit holds the output of our proposed DMR voter till the fault effect disappears. Proposed low-cost HFS-box provide a high capability of fault tolerant against transient faults with any duration by putting low area overhead, i.e. 137%, and low throughput degradation, i.e. 11.3%, on the original implementation.

## I. Introduction

Dependable applications, like secure information systems, remote security services, online banking, etc play an important role in our daily lives. Secure storage and communication are critical requirements of these applications. Nowadays, cryptography is extensively used in dependable applications to meet these critical requirements and in consequence, prevent any unauthorized access to the secure information. Another important requirement of dependable application is reliability. Therefore, in many cases, a fault resilient approach incorporated with original hardware implementation [1].

The Advanced Encryption Standard (AES) [2] was standardized by the National Institute of Standards and Technology (NIST) in 1997. After that AES has been one of the most common symmetric cryptographic algorithms. Until now, many hardware implementations of the AES were proposed with different characteristics [3-6] which each of them is suited for different applications with different constraints. Recently, many fault injection attacks has been proposed on AES [7-9]. In a fault attack, attackers inject malicious faults into the VLSI design of cryptographic primitives to extract secure information (i.e. cryptographic key).

On the other hand, with transistor size downscaling, reducing power supply voltage level, increasing operating frequencies and therefore reducing noise margins, VLSI hardware designs will be more and more sensitive to random faults occurrence [10]. All random faults that occur in VLSI designs can be grouped into transient and permanent faults.

To thwart the random and/or malicious faults effect, various fault resilient hardware implementations of AES were proposed [11-14]. AES include four basic operations, named SubByte, ShifRows, MixColumns and AddRoundKey. The hardware implementation of SubByte operation is realized with 16 S-Boxes that are nonlinear mapping in which replace each byte of state array with another byte. It also occupies much of the total AES hardware implementation area [15]. So, integrating its hardware implementations with an efficient fault resilient scheme is crucial for making the AES robust to the random and/or malicious faults. There are many online error detection schemes for SubByte implementation of AES, see for example [16-17].

Just a few studies among previous research works have addressed the fault correction. In fact, in most of the previous studies only detection task is considered and so, for their solutions the extra corrective operations should be employed. In [18] a hybrid redundancy in which hardware redundancy and time redundancy are combined for fault correction in S-box is proposed. Their proposed S-box architecture can tolerate the single faults. It's worth noting that the fault tolerant S-box in [18] provided a high level of reliability against the natural faults due to the essence of electronic devices, not the malicious faults in the fault attacks. The main aim of the present paper is to propose a high-throughput fault-resilient hardware implementation of the AES S-box. We propose a correction scheme in hardware level so that the circuit frequency is not significantly affected. In this paper a high-speed design is considered. In fact, we exploit the features of gate-level implementation of S-box allowing pipeline technique to speed up the hardware implementation of SubByte operation of AES. The proposed technique is also practical for any generic cipher block.

The main contributions of this paper are including as follows:
- We present an implementation of high-throughput and lightweight S-box in the gate level for high-speed AES encryption.
- We propose a fault-resilient technique, i.e. FC-DMR, for real-time applications which cannot tolerate high running time and require a high-speed process. Proposed technique generally could be used in all digital functional units.
- We design a new DMR voter which is composed of the standard library components and could be implemented on any digital platform such as FPGA and ASIC.

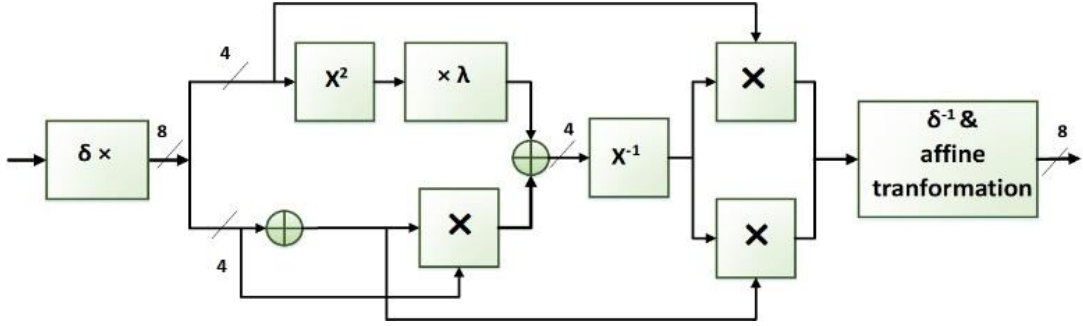

Fig. 1. Composite field based S-box architecture.

The rest of this paper is organized as follows. Section II presents a brief background of the S-box of AES algorithm and its implementation. Section III presents the proposed fault-resilient technique (FC-DMR) besides our DMR voter model.

It also describes the HFS-box architecture. We evaluate the proposed architecture's architectural characteristics in terms of area, frequency and throughput in section IV. Finally, section V concludes the paper.

## II. S-BOX IMPLEMENTATION

In this subsection, we describe the S-box operation and its utilized architecture. The proposed S-box architecture using composite-field in [19] is employed in this paper. The S-box operation which is believed to be most resource consuming among other AES operation, is a nonlinear mapping on each state array byte. This nonlinear mapping is nothing but finding a multiplicative inverse over $GF(2^8)$, i.e. $x^{-1} \epsilon\, GF(2^8)$ followed by an affine transformation. In other words, if $y = SB(x)$ and $X \epsilon GF(2^8)$ and $Y \epsilon GF(2^8)$, then we have:

$$y = Ax^{-1} + b = \begin{bmatrix} 1 & 1 & 0 & 0 & 0 & 0 & 1 & 0 \\ 0 & 1 & 0 & 0 & 1 & 0 & 1 & 0 \\ 0 & 1 & 1 & 1 & 1 & 0 & 0 & 1 \\ 0 & 1 & 1 & 0 & 0 & 0 & 1 & 1 \\ 0 & 1 & 1 & 1 & 0 & 1 & 0 & 1 \\ 0 & 0 & 1 & 1 & 0 & 1 & 0 & 1 \\ 0 & 1 & 1 & 1 & 1 & 0 & 1 & 1 \\ 0 & 0 & 0 & 0 & 1 & 0 & 1 \end{bmatrix} x^{-1} + \begin{bmatrix} 0 \\ 0 \\ 0 \\ 0 \\ 1 \\ 0 \\ 1 \\ 1 \end{bmatrix} \quad (1)$$

Since direct multiplicative inversion of S-box computation is costly, multiplicative inversion in composite fields is preferred [20]. This implementation leads to lower complexity and smaller implementation area. The S-box implementation using composite-field and polynomial basis is illustrated in Fig. 1. As shown in this figure, the 8-bit input of multiplicative inversion, i.e., $X = \sum_{i=0}^{7} \alpha_i x_i$ in the binary field $GF(2^8)$, using the transformation matrix δ transforms to composite-field $GF(2^8)/GF(((2^2)^2)^2)$. In turn, the output of the multiplicative inverse from composite-field transforms back to binary field $GF(2^8)$ by the inverse transformation matrix $\delta^{-1}$ to obtain $X^{-1}$. The hierarchical composite-field decomposition, i.e., $GF(((2^2)^2)^2) \to GF((2^2)^2)$, $GF((2^2)^2) \to GF(2^2)$ and $GF(2^2) \to GF(2)$, can be made using the irreducible polynomials of $x^2 + x + \lambda$, $x^2 + x + \varphi$ and $x^2 + x + 1$, respectively. As shown in Fig. 1, the output of S-box i.e., Y, is obtained using the affine transformation after inverse transformation ($\delta^{-1}$) [19]. The S-box compose of the multiplications, the squaring and the inversion that all of them are over $GF((2^2)^2)$. Besides these arithmetic blocks, the S-box includes modulo-2 addition that realized by XOR gates, see Fig. 1. Considering this figure, the output of the S-box can be formulated as following:

$$\sigma_h = ((\xi_h + \xi_l)\xi_l + \xi_h^2 \lambda)^{-1} \xi_h \quad (2)$$
$$\sigma_l = ((\xi_h + \xi_l)\xi_l + \xi_h^2 \lambda)^{-1} (\xi_h + \xi_l) \quad (3)$$

Where, the $\xi$ and $\sigma$ are the input and output of the multiplicative inversion, respectively.

## III. PROPOSED FAULT CORRECTION STRUCTURE (FC-DMR)

### A. FC-DMR

We propose a correction technique in a DMR implementation of a digital circuit (FC-DMR) depicted in Fig. 2. The proposed FC-DMR protects the operation of both combinational and the sequential parts of a digital circuit in each pipeline stage. Fig. 2 depicts an instance pipeline stage i in the intended circuit. As depicted in this Figure, our FC-DMR is consist of the following elements:

- *Pipeline Logic$_i$ (original)*: a part of system's combinational logic utilized to process data in the original mode in the i$^{th}$ pipeline stage.
- *Pipeline Logic$_i$ (redundant)*: a redundant copy of the original i$^{th}$ pipeline stage utilized to process data in the redundant mode in the i$^{th}$ pipeline stage.
- *Register stage$_i$*: the register or sequential part of the i$^{th}$ pipeline includes DMR register and two DMR voters to preserve the correct state in present of fault.
- *DU*: the fault detection unit which is actually implemented using a comparator must provide the output error signal *err$^i$* which indicate any differences in the DMR register in the i$^{th}$ pipeline stage occur.
- *CU*: the control unit produce the *Err* which is a general error signal and indicates a fault occurrence in the system (any pipeline stage), i.e., a fault is detected.

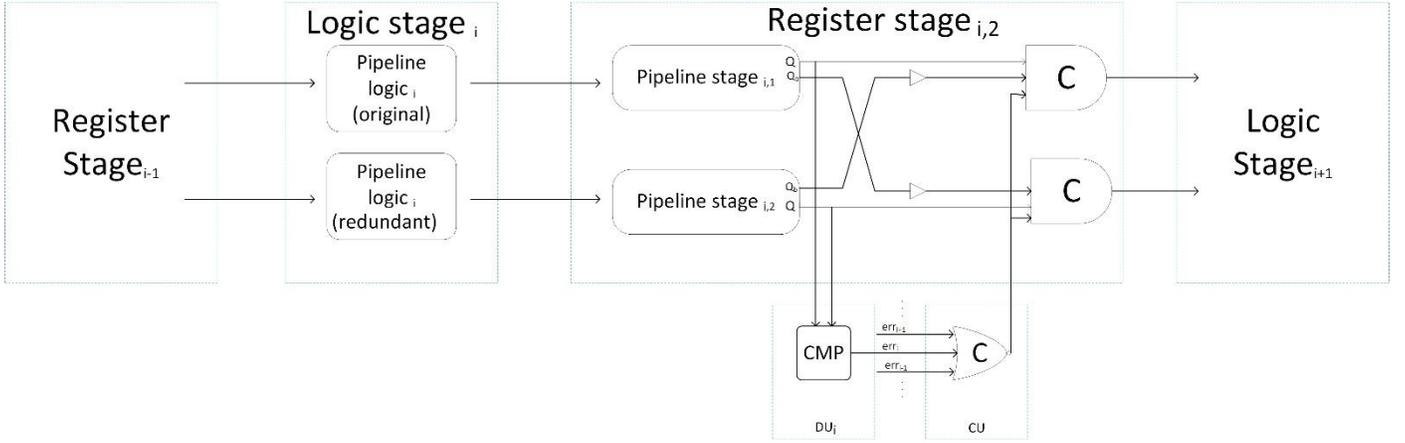

Fig. 2. Proposed fault correction technique in DMR implementation (FC-DMR).

The input of each pipeline stage is processed by the pipeline logic and its redundant unit. The corresponding output of the original and redundant pipeline logic units are stored in the register stages, i.e. pipeline register$_1$ and pipeline register$_2$, respectively. If the register's contents are identical, no fault(s) is (are) detected. Otherwise, the comparator $CMP^i$'s output in $DU^i$, i.e. $err^i$, will be activated. Two DMR voter are employed to protect both combinational and sequential part of the system. Proposed technique can correct any transient fault which occurs in a single S-box. When a fault detects in any pipeline stage components, either in the logic stages, in the pipeline registers, or in the $DU$, the $CU$ will reset its output, i.e. $Err$ and later it prevents loading the incorrect state on the output of DMR voters.

Hence, the pipeline logics process previous correct state till the fault effect disappears. When the fault effect disappears, the next correct state will process without any problem. This solution may put a negligible delay overhead on the critical path due to the comparison and voting.

### B. Proposed voter

The employed voter does the two tasks of a majority voter in a DMR technique which is: holding the previous state when faces a mismatch and changing the vote signal's value when both modules produce a same output.

In fact, when the outputs of the two replicas are not the same as each other which means an error has occurred, the voter holds the previous value until the two replicas' outputs become similar. Besides, our design has a delay module which is useful in case of the comparator faces a mismatch. This delay makes it possible to affecting the enable signals. Enables are provided to control internal wires not to send the faulty signals to voter's output which means that pipeline stage be unchanged until the correct value gains and the sequence in our pipeline design remains unaffected.

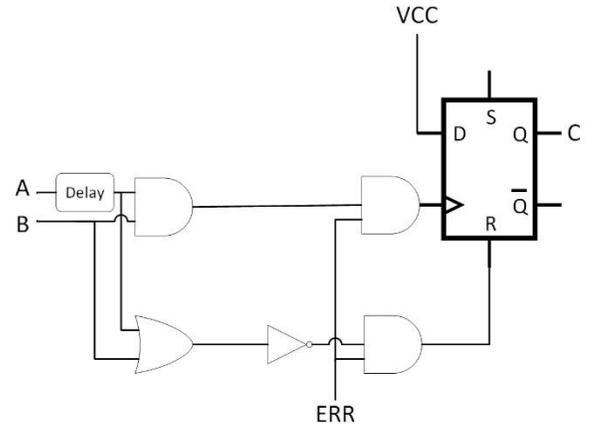

Fig. 3. Proposed voter in gate level.

### C. HFS-box

The main contribution of this paper is proposing high-throughput fault-resilient hardware implementations of S-box.

We propose a fully pipeline implementation of S-box in composite field approach which leads to reduction of the circuit critical path. In fact, this solution enables us to enhance frequency of clock signal in our proposed method and also makes it suitable for meeting the high-speed application requirements.

Proposed pipeline S-box is depicted in Fig. 4. We place pipeline registers into this schema which are illustrated by the dotted lines. As depicted in this figure, proposed S-box architecture (shown in Fig. 1) is divided into 5 stage. This pipeline registers are inserted to S-box architecture so that the critical path is optimally pipelined. This architecture is integrated with proposed FC-DMR to achieve fault tolerance for any transient fault in both combinational and sequential parts in any pipeline stage of a single S-box, named HFS-box. In HFS-box each DMR implementation of pipeline logic is lied between two register stages to check against fault occurrence as depicted in Fig. 2.

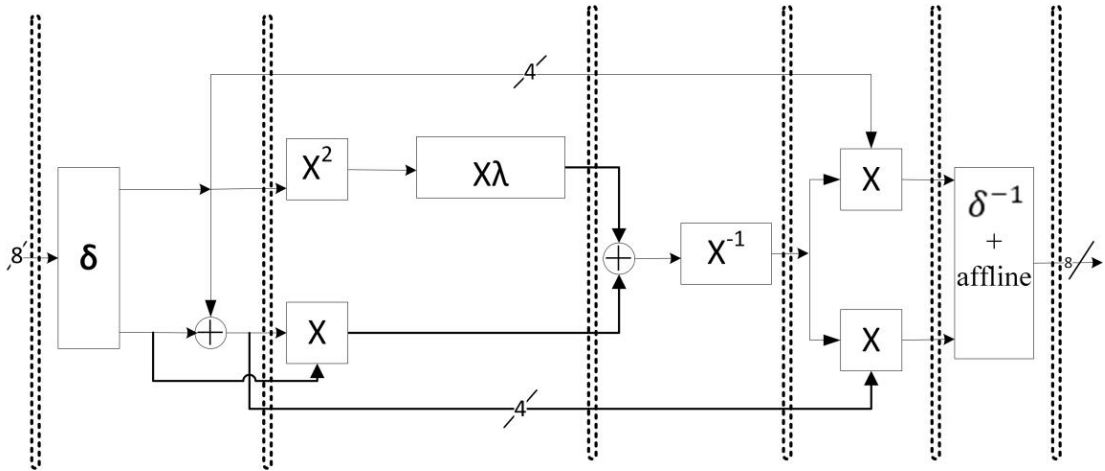

Fig. 4. The architecture of the S-box with 5-stage pipeline.

## IV. IMPLEMENTATION RESULT

To evaluate the proposed HFS-box, we compare it with the TMR and TTR implementation of S-box, as traditional fault tolerant structures with high fault correction capability. We report the synthesis result by using the TSMC 180 nm CMOS. We employ Verilog as design entry description language and Synopsys DC as the-synthesis tool. It should be noted the 8-bit SubByte operation is considered so in each structure a single S-box is needed.

Table 1. Throughput, maximum frequency, area result.

| Design metric | | Original | TMR | TTR | HFS-box |
|---|---|---|---|---|---|
| Area | GE | 212.42 | 673.31 | 279.02 | 503.46 |
| | % | - | 216 | 31.35 | 137 |
| Frequency | MHz | 555 | 525 | 519 | 492 |
| | % | - | -5.4 | -6.4 | -11.3 |
| Throu. | Mbps | 4440 | 4200 | 1384 | 3936 |
| | % | - | -5.4 | -68.8 | -11.3 |
| Fault tolerance | Transient | ✗ | ✓ | ✓ | ✓ |
| | Permanent | ✗ | ✓ | ✗ | ✗ |
| Security against fault attack | | ✗ | ✓✗ | ✓✗✗ | ✓ |

In this section, the ASIC implementation results of all fault-tolerant S-box implementations are reported and compared. The design features that we consider contain area, area overhead, frequency and frequency overhead. Table 1 presents the implementation results of all fault resilient designs.

In this table, we use equations 4 to calculate the the cost overhead.

$$Overhead = \frac{C_{FT} - C_O}{C_O} \quad (4)$$

Where, $C_O$ is the original implementation cost (area, frequency, throughput, etc.), and $C_{FT}$ is the cost of the fault tolerant implementation. It can be seen that TTR has the lowest area overhead (44.5% and 58.54% reduction compared to HFS-box and TMR, respectively) and at the same time lower throughput, (64.83% and 67.04% worse than HFS-box and TMR, respectively). HFS-box requires about 503 NAND gate equivalences (GEs). Actually, it puts more area overhead than TTR but still is much better than TMR (25.22% better than TMR). However, TMR achieves the best throughput among all fault resilient architectures, its security and reliability against fault attacks is lower than our HFS-box and also it puts much more area overhead on the original S-box than HFS-box. In fact, proposed low-cost HFS-box can continue its proper task without considerable negative impact on the system speed or even any traditional recovery scheme. It is a suitable fault tolerant technique for resource-constrained applications that require a high level of security.

## V. CONCLUSION

In this paper, we proposed a lightweight high-throughput fault-resilient architecture for composite field S-box implementation of AES which consume the largest space in AES, named HFS-box. The proposed fault-resilient technique is based on fault correction in DMR implementation (FC-DMR) combined with a temporal redundancy technique. It is able to correct transient faults which may occur in S-box naturally or maliciously. Our solution is valid for any digital circuit implementation (specially block cipher hardware implementation) with different level of pipelining. HFS-box uses 5 pipeline stage to meet the real-time application requirements for speed and throughput. Indeed, we inserted pipeline registers in optimal places in the S-box architecture. Furthermore, we introduced a compatible DMR voter with our FC-DMR. The proposed HFS-box and two well-known methods with high fault-tolerant ability, i.e. TMR, TTR have been implemented on ASIC using TSMC 180nm CMOS technology and their area, frequency and throughput have been derived and reported. The synthesis results pointed out that the HFS-box has a low area overhead (137%) and low throughput degradation (11.3) compared with other fault tolerant schemes.